\documentclass[aps,preprint]{revtex4-1}

\pdfoutput=1

\usepackage[utf8]{inputenc}
\usepackage{lmodern}
\usepackage{hyperref}
\usepackage{amsmath} 
\usepackage{graphicx}
\usepackage{amssymb}
\usepackage{amsthm}
\usepackage{epstopdf} 
\usepackage{natbib}
\usepackage{dcolumn}
\usepackage{amsfonts}

\newcommand{\cmc}{\,\mathrm{cm}^{-3}}
\newcommand{\gvm}{\,\mathrm{GV/m}}

\newcommand{\wcm}{\,\mathrm{W/cm}^2}
\newcommand{\mic}{\,\mu\mathrm{m}}
\newcommand{\MeV}{\,\mathrm{MeV}}
\newcommand{\fs}{\,\mathrm{fs}}

\newcommand{\pc}{\,\mathrm{pC}}
\newcommand{\fc}{\,\mathrm{fC}}
\newcommand{\mrad}{\,\mathrm{mrad}}

\begin{document}

\title{Concept of a laser-plasma based electron source for sub-10 fs electron diffraction}

\author{J. Faure, B. van der Geer, B. Beaurepaire, G. Gall\'e, A. Vernier and A. Lifschitz  }

\affiliation{Laboratoire d'Optique Appliqu\'ee, UMR 7639 ENSTA-CNRS-Ecole Polytechnique, 91761 Palaiseau, France}%

\begin{abstract}
We propose a new concept of an electron source for ultrafast electron diffraction with sub-10~fs temporal resolution. Electrons are generated in a laser-plasma accelerator, able to deliver femtosecond electron bunches at 5 MeV energy with kHz repetition rate. The possibility of producing this electron source is demonstrated using Particle-In-Cell simulations. We then use particle tracking simulations to show that this electron beam can be transported and manipulated in a realistic beamline, in order to reach parameters suitable for electron diffraction. The beamline consists of realistic static magnetic optics and introduces no temporal jitter. We demonstrate numerically that electron bunches with 5~fs duration and containing 1.5~fC per bunch can be produced, with a transverse coherence length exceeding 2~nm, as required for electron diffraction.
\end{abstract}

\maketitle

\section{Introduction}
In the last decade, there has been great progress in plasma-based techniques for accelerating particles \cite{esar09}. In particular, in laser-plasma accelerators, an ultra intense laser pulse drives a relativistic electron plasma wave - or wakefield - in the plasma. This wakefield is equivalent to an accelerating structure with extremely high accelerating gradients, on the order of $100\gvm$, and is able to accelerate electrons to relativistic energies in micrometer distances. Currently, laser-plasma accelerators provide electron beams ranging from hundreds of MeV \cite{faur04,gedd04,mang04} to multi-GeV energies \cite{wang13,leem14} and charges in the tens of picoCoulomb range. Numerous methods have been developed to control the injection of electrons into the wakefield \cite{faur06,gedd08,schm10,mcgu10,pak10}, resulting in energy spreads at the few percent level \cite{rech09b}. In addition, when the injection is done properly, femtosecond electron bunches can be generated: for example, in Ref.~\cite{lund11}, a 80~MeV electron beam with 1.5~fs r.m.s. duration was experimentally demonstrated and measured. This emerging technology holds the promise of compact particle accelerators delivering high charge femtosecond bunches with intrinsic synchronization to an optical pulse. 
Jitter-free synchronization originates from the fact that the accelerating structure is directly driven by the laser. Therefore, synchronization and temporal jitter issues should not be of concern and sub-10~fs temporal resolution should, in principle, be attainable in experiments.

These unique properties make laser-plasma accelerators attractive for applications requiring femtosecond bunches and high peak currents, such as the development of compact Free Electron lasers (FEL). Several groups are now engaged in the demonstration of a FEL using a laser-plasma source \cite{maie12,huan12,loul15}. FELs in the X-ray energy range have been demonstrated recently - at the LCLS for example  - in very large infrastructures based on conventional Radio-Frequency (RF) accelerator technology \cite{emma10,ishi12}. Such machines produce bright coherent X-ray flashes with femtosecond duration \cite{grur12} which are used by a large variety of users to investigate the ultrafast dynamics of matter at the femtosecond time scale using pump-probe experiments \cite{miller14}. 

Recently, ultrafast electron diffraction (UED) \cite{miller14} has also emerged as an alternative technique for studying ultrafast structural changes in solids, gases or liquids \cite{zewail09,scia11}. Although electron bunches might not be as versatile as X-rays, they provide a powerful structural probe while the required experimental set-ups are very compact and relatively low cost. However, the time resolution in UED experiments has been limited to $> 100-200$~fs, either because of space charge \cite{scia11} or RF-jitter \cite{oudh10,musumeci10}. Reducing the temporal resolution to the sub-100 fs range and even the sub-10 fs range, remains a challenge of importance as it could give access to the observation of new phenomena and new physics in pump-probe experiments.
 
 In this article, we use numerical simulations to show that a laser-plasma accelerator can provide an electron source well suited for UED with sub-10 fs resolution. The laser-plasma accelerator we have designed is able to produce pC of charges in the 5 MeV range and with a $< 10$~fs bunch duration. In section \ref{secPIC}, we use Particle-In-Cell (PIC) simulations to show that such bunches can be generated at kHz repetition rate. In section \ref{secBeamLine}, we tackle the problem of beam transport of the electron source: we design a beamline whose goal is to produce a high quality beam (small transverse emittance, narrow energy spread) with sub-10 fs duration at the sample position, tens of centimeters downstream the electron source. This problem is particularly challenging because it is rather difficult to transport electron beams from a plasma accelerator as their energy spread and divergence are larger than in conventional accelerators. Therefore, maintaining the emittance and the femtosecond bunch duration is not trivial. Beam transport of laser-driven electron beams has been studied in the context of a driver for a FEL \cite{maie12,loul15}, but not as a driver for a UED experiment. Finally, section \ref{secGPT} shows results of the beamline optimization using the General Particle Tracer (GPT) code \cite{GPT}. We show that the beamline is able to provide $\simeq 5$~fs bunches with femtoCoulomb charge and normalized transverse emittances of $\varepsilon_{n\bot}\simeq10-20$~nm.

\section{Laser-plasma electron source}\label{secPIC}

We start with the description of the production of the electron beam. Electrons are produced during the interaction of a ultra-intense and ultrashort laser pulse with an underdense plasma. The ultra-intense laser pulse perturbs the electron density via the ponderomotive force and excites a large amplitude plasma wave, in which electrons can be trapped. The maximum energy that electrons can gain in this process is given by $\Delta E\propto mc^2\gamma_p^2\delta n/n_0$, where $m$ is the electron mass, $c$ is the velocity of light, $\delta n/n_0$ the amplitude of the plasma wave and $\gamma_p=(n_c/n_0)^{1/2}$ is the Lorentz factor corresponding to the plasma wave phase velocity $v_p$, with $n_c$ and $n_0$ respectively the critical density and the electron plasma density. Therefore, at high density $n_0/n_c\simeq 0.1$ (i.e. $n_0\simeq 10^{20}\cmc$), plasma accelerators can produce electrons in the $10$~MeV range. 

This energy range is suitable for UED: electron diffraction in the $1-10$~MeV range has been demonstrated experimentally and can produce high quality diffraction patterns \cite{hastings06,murooka11,musumeci10b}. However, using electrons with energies larger than $10$~MeV is very challenging because of the diffraction angles: $\sin\theta=\lambda_{dB}/2d_{hkl}$ where $\lambda_{dB}$ is the de Broglie wavelength of the electrons and $d_{hkl}$ is the interplane spacing. At high energy, $\theta$ becomes too small and obtaining the diffraction pattern becomes cumbersome. In addition, at $E>10$~MeV, Bremsstrahlung radiation can become prominent and could perturb the diffraction measurements. 

Thus, we aim at producing electrons with femtosecond durations and with energies smaller than $10$~MeV. Previous experiments in the self-modulated laser wakefield have shown that using a 30~fs, high energy laser pulse ($\simeq 1$~J) in a high density plasma can lead to copious amounts of electrons in the $1-10$~MeV range \cite{malk01}. Electrons were self-injected \cite{malk01} or injected through ionization \cite{guil15} and the obtained bunches did not have femtosecond durations, but rather tens of femtoseconds. In addition, the beam quality was rather poor with a large energy spread and a large divergence. Although operating the laser-plasma accelerator in this regime is possible, it is well known that higher quality bunches can be generated in the \emph{bubble regime} \cite{pukh02}, which is obtained when the laser pulse length is matched to the plasma wavelength. In this regime, the laser pulse ponderomotive force drives a nonlinear plasma wave in which electrons form a spherical electron sheath around the ions (the so-called plasma bubble). In this case, electrons can be trapped into this spherical bubble, leading to a divergence in the range of a few mrad, an energy spread of $1-10\%$ and a duration of a few femtoseconds \cite{faur04,lund11}.  

\begin{figure}[th!]
\centerline{\includegraphics[width=8.5cm]{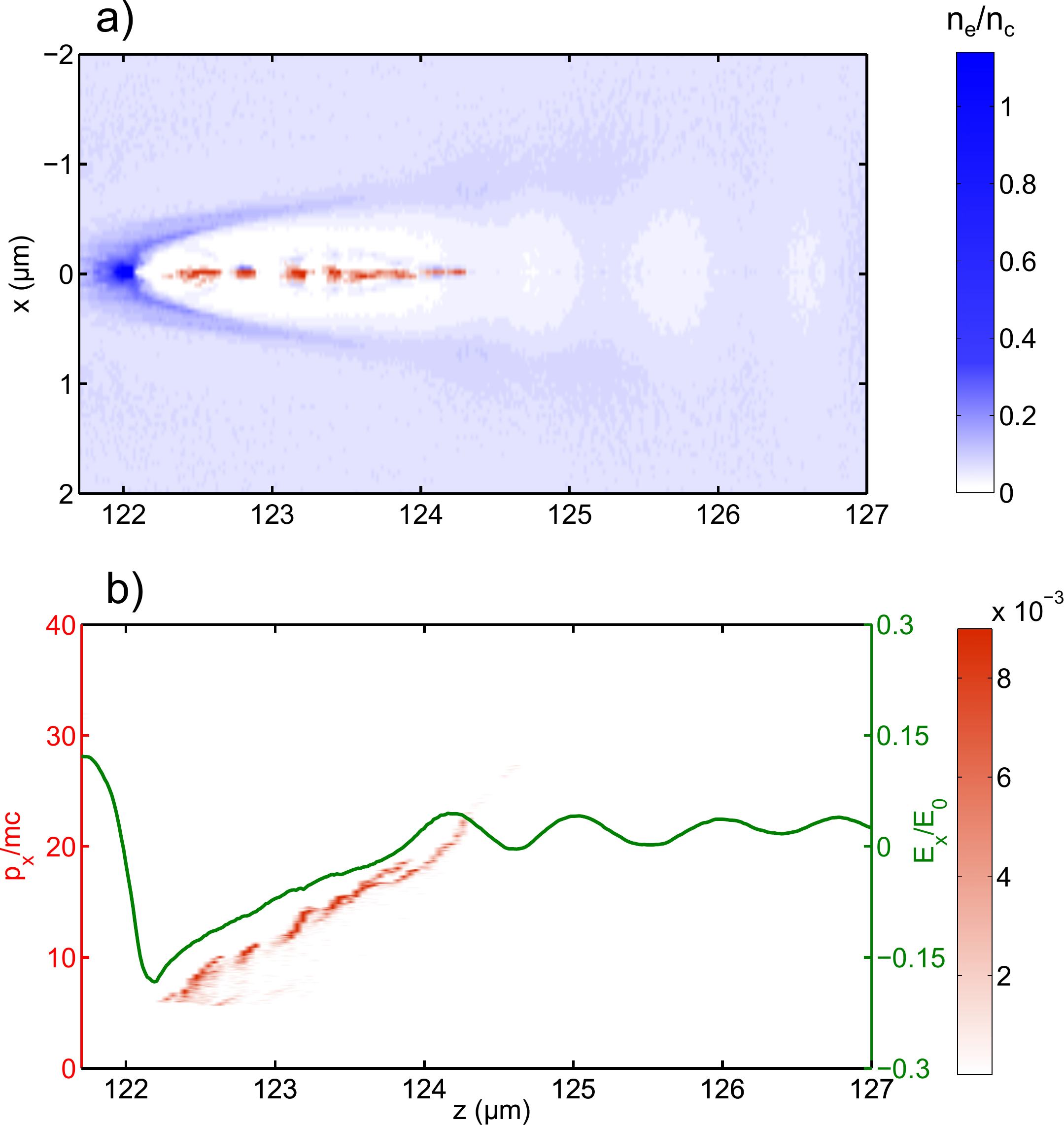}}
\caption{Results of PIC simulations. a) Map of the electron density in the plasma ($n_e/n_c$ blue color bar): the 5~fs laser pulse drives a plasma cavity, in which high energy electrons are accelerated (red color bar). b) Longitudinal phase space density map showing an accelerated electron bunch with energy in the 5-10 MeV range. The green curve is the wakefield longitudinal electric field originating from the plasma cavity. }\label{fig0}
\end{figure}

The bubble regime can be obtained for a given set of laser and plasma parameters \cite{lu07}, the basic idea being that the laser pulse is matched longitudinally  and transversely to the plasma wave. Assuming that $R$ is the size of the spherical bubble, the longitudinal matching condition can be written as $c\tau\simeq R/2$,  where $\tau$ is the laser pulse duration. The transverse matching conditions reads $k_pR=2\sqrt{a_0}$, where $a_0=eA_0/mc\omega_0 $ its normalized vector potential, and $k_p=\omega_p/c$ is the plasma wave vector. From these simple considerations, one can derive scaling laws \cite{lu07} showing that to accelerate electrons in a plasma bubble in the $10$~MeV range, one needs a $5$~fs laser pulse with $\simeq 5$~mJ, focused to a $w_0=2.5\mic$ waist. 

Few milliJoules and few-cycle laser pulses are currently obtained in state-of-the-art kHz laser systems \cite{bohm10,bohle14}. The high-repetition rate of the mJ lasers is a great advantage for pump-probe experiments compared to Joule-level laser systems operating at 10 Hz or less. Indeed, UED requires the detection of changes at the percent level in the intensity of the Bragg peaks and averaging data at kHz helps to increase the statistics and wash out the charge fluctuations inherent to plasma accelerators. Note that laser-plasma accelerators operating at kHz repetition rate were recently demonstrated \cite{he13} as well as their application to electron diffraction \cite{he13b}. Higher stability was observed in these high repetition rate experiments. However, the electrons were accelerated in the 100 keV range and the bunch length quickly stretched due to velocity dispersion. The experimental demonstration of a kHz laser-plasma accelerator delivering MeV and femtosecond electron bunches is still pending.

 \begin{figure*}[th!]
\centerline{\includegraphics[width=16cm]{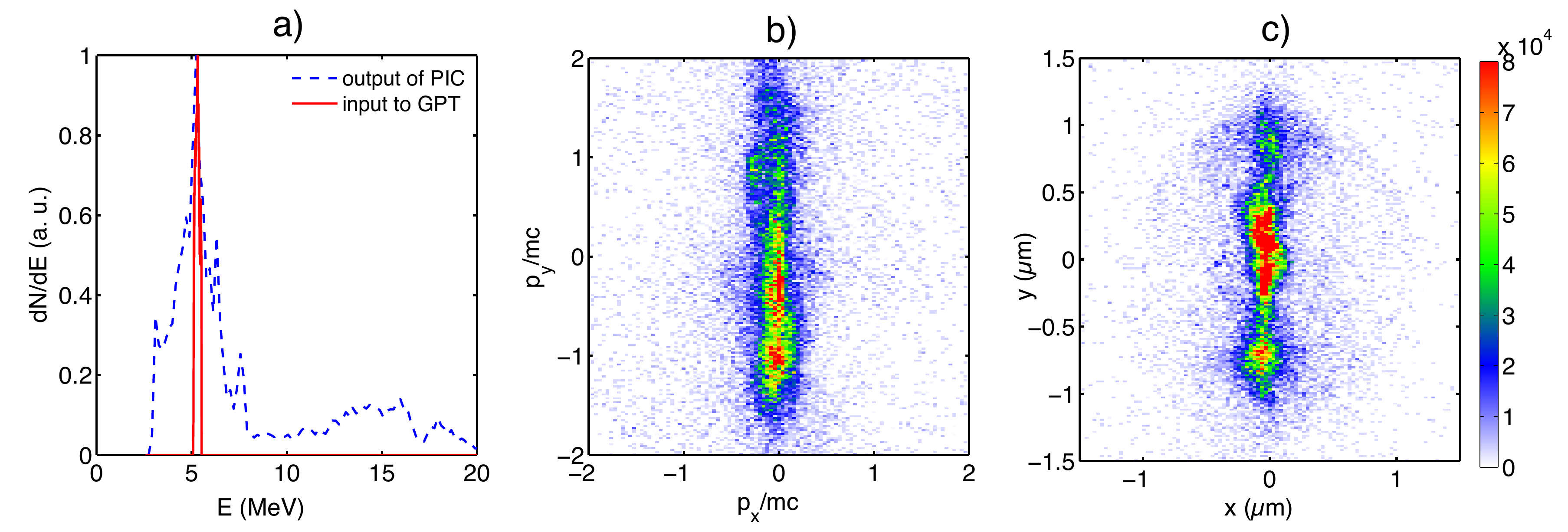}}
\caption{Results from PIC simulations. a) blue line: electron energy distribution at the output of the PIC code. Red line: the filtered energy distribution ($\delta E/E=2\%$) which is used as an input of the particle transport code (GPT). b) Electron distribution in transverse phase space $(p_x,p_y)$ and c) in real laboratory space $(x,y)$. Both plots show an elongation of the distribution along the direction of laser polarization. }\label{fig1}
\end{figure*}

In order to investigate precisely the parameters of the electron beam in this regime, we simulated laser–plasma interaction using the fully electromagnetic PIC code Calder–Circ which provides a quasi-3D geometry \cite{lifs09}. We ran simulations with realistic experimental parameters, namely a 5~fs full width at half maximum (FWHM) pulse centered around $\lambda_0=800$~nm with an energy of $4.1\,$mJ. The pulse is propagating in an underdense plasma, whose longitudinal density profile consists of a $203\mic$ plateau preceded and followed by a $63\mic$ ramp. The electronic density on the plateau is $n_0 = 0.05 \times n_c = 8.7 \times 10^{19}\cmc$. The laser beam is focused at the beginning of the plateau, down to a waist of $4.3\mic$, so that the laser intensity is $I = 2.6 \times 10^{18}\wcm$ - the corresponding value of the normalized vector potential is $a_0 = 1.1$. The laser is polarized along the $y$-axis and propagates along the $z$-axis. The box is composed of $1500 \times 200$ cells with 220 particles per cell. The longitudinal cell size is $0.125c/\omega_0$ and the radial cell size is $0.628c/\omega_0$.

The physics of this interaction has been studied in details in a previous publication \cite{beau14}. We showed that the highly nonlinear evolution of the laser pulse causes the slow down of the wakefield whose phase velocity becomes sub-relativistic. Electrons are then injected in this ``slow'' wakefield and accelerated to energies in the 5-10~MeV range. Figure \ref{fig0}a) shows the electron density in the plasma. One can clearly see the plasma bubble (blue color bar) and the electron bunch (in red) which has been injected and accelerated into it. In Figure \ref{fig0}b), the electron bunch is represented in the phase space $(z,p_z)$. Clearly, electrons have been accelerated at the $5-10$~MeV level, and the bunch spreads over $2\mic$, i.e. $\simeq 6$~fs.

Figure \ref{fig1}a) shows the electron energy distribution when the electron beam exits the plasma. The distribution (dashed blue line) has a relatively large peak at  $5$~MeV ($\delta E/E=23\%$ at FWHM) and contains $7\pc$ of charge. The red line shows the distribution which was trimmed to a smaller energy spread $\delta E/E=2\%$ FWHM and still contains a charge of $\sim1\pc$ (this trimmed distribution is used for the design of the beam transport system in the following section). Figure \ref{fig1}b) shows the electron distribution in transverse phase space $(p_x,p_y)$. The transverse momentum is much larger in the direction of the laser polarization ($y$-direction) indicating the electron beam has gained more momentum in this direction because of its direct interaction with the laser field. A similar behavior is seen in figure \ref{fig1}c) which represent the beam distribution in $(x,y)$: the beam is more elongated along the $y$-axis. Note the very small source size, $\sigma_x\simeq 0.1\mic$ in the $x$-direction. The normalized r.m.s. emittance of this beam is $\varepsilon_x=10$~nm and $\varepsilon_y=300$~nm (including the useful part of the charge, i.e. $75\%$ of the total charge, see table \ref{table1} for more details). Even though the transverse emittance is good, the beam has a large divergence $\theta_x=15\mrad$ and $\theta_y=80\mrad$.

This laser-plasma accelerator is able to provide a beam at  $5\MeV$, with a charge density of $\simeq 500\fc/$ per \% of $\delta E/E$ , a relatively good transverse emittance and most importantly, a sub-10~fs bunch duration.

\section{Beamline design}\label{secBeamLine}

The electron beam described in the previous section cannot be used directly for electron diffraction for several reasons. First, the diffraction sample has to be placed sufficiently far away from the source so that it does not get damaged by the laser pulse which generates the electron beam. This implies that the electron beam has to propagate tens of centimeters before reaching the sample. Since the energy spread is quite large, the bunch duration increases upon a vacuum drift of length $L_{d}$. For a relativistic beam with $\gamma^2\gg1$, the bunch stretches by $dt$: 
\begin{equation}
dt=\frac{L_{d}}{c}\frac{\delta \gamma}{\gamma^3}
\end{equation}
For example, after a $20$~cm propagation, a $5$~MeV bunch with an energy spread of $2\%$ will stretch by $130\fs$. Similarly, energy fluctuations of $2\%$ will translate to a temporal jitter of $130\fs$. The bunch emerging from the laser plasma interaction has a $20\%$ energy spread and is likely to fluctuate in energy. It is clear that in order to keep the advantages of using the initially sub-10 fs bunches, it is necessary to design a beamline which will satisfy the following conditions: (i) restrict the energy spread to the $1\%$  level, (ii) fix the absolute energy in order to avoid energy fluctuation which would degrade the temporal resolution by introducing a temporal jitter, (iii) recompress the electron bunch duration at a given position, tens of centimeters away from the source.

 \begin{figure}[t!]
\centerline{\includegraphics[width=8.5cm]{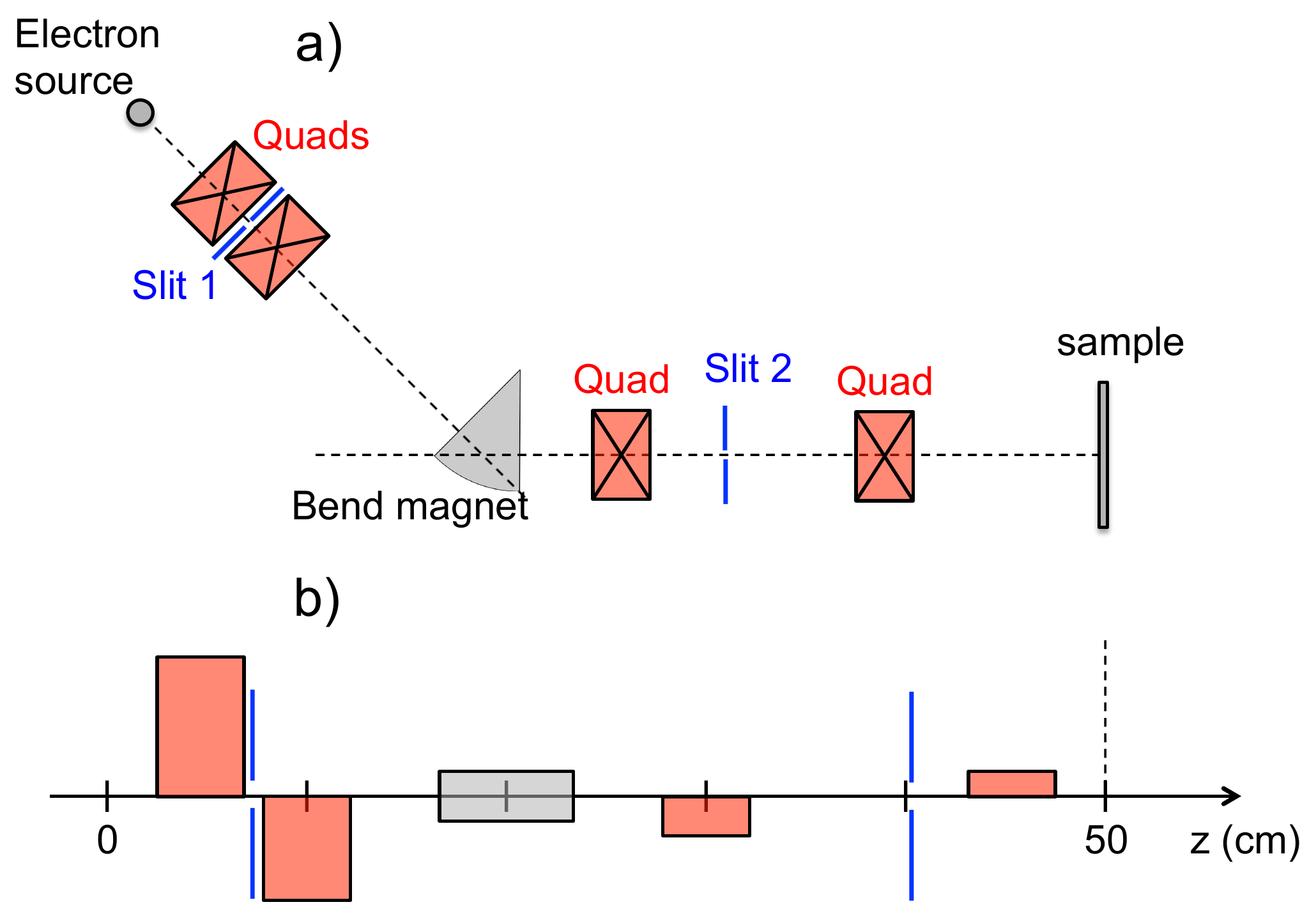}}
\caption{a) Schematic representation of the static compression beamline. Note the presence of two slits which are used to trim the electron distribution. b) Conventional representation of the beam line showing distance between magnetic elements. The sample plane lies at $z=50$~cm.}\label{fig2}
\end{figure}

In order to obtain high quality diffraction images, the beam should also have a very good transverse emittance. Electron diffraction is a single electron phenomenon: constructive interference in the direction of the Bragg angles will only occur if the electron wave packet has a large transverse coherence $L_{cx}$ (along $x$). The transverse coherence is related to the beam transverse normalized emittance by
\begin{equation}
L_{cx}=\frac{\hbar}{mc}\frac{\sigma_x}{\epsilon_{nx}}
\end{equation}
where $\sigma_x$ is the r.m.s. beam size along $x$. Typically, $L_{cx}$ should be larger than the size of the unit cell of the crystal under consideration. For instance, taking $L_{cx}>1$~nm and $\sigma_x=100\mic$ implies a very good emittance: $\varepsilon_{nx}<30$~nm. Therefore the beamline should be able to maintain the beam emittance while providing the correct beam size, $\sim 100\mic$ on the sample.

Finally, there are other design considerations that need to be taken into account: (i) we would like the whole beamline to be compact, $<1$~m and to contain only a limited number of elements in order to reduce its cost and complexity; (ii) the beamline should no use RF elements in order to avoid time jitter which would spoil the temporal resolution.

Such a beamline is not standard, especially when considering the sub-10~fs target for the bunch duration. In particular, it is desirable that the compression be performed using a magnetic bunch compressor in order to avoid temporal jitter. In Ref.~\cite{tokita09} a beamline was used for compressing a laser-driven electron bunch, but the final duration was of several hundreds of femtoseconds. Chicanes are used in large machines for shaping and compressing electron bunches but they contain a large number of bend magnets and quadrupoles \cite{emma10}. Alpha-magnets are also known to compress particle bunches \cite{kung94}, however we found that it was impossible to use an alpha-magnet for compressing the bunch in less than a meter. Therefore, we have chosen to use a simple bend magnet. The idea of the beamline is presented in fig.~\ref{fig2}: the first pair of quads is used to focus the beam into the bend magnet, while the second pair of quads is used to adjust the beam size in the sample plane. The $45^{\circ}$ bend magnet provides compression of the bunch but also disperses it in the $x$-plane so that using a slit, the central energy and the energy spread can be tuned. The first slit is used to trim the angular distribution while the second restricts the beam size and the energy spread. Note that the whole beamline fits within 50~cm.

For such short bunches, there are large spatio-temporal couplings that appear during propagation in the beam line. Electrons travelling with larger angles fall at the back of the bunch and we find that it is very challenging to recompress the bunch when the electron source has a large divergence angle. In order to gain insight on bunch dynamics in the beamline, we used a ray tracing model where magnetic optics are described in the framework of the hard edge approximation \cite{AccPhys99}. The initial conditions of an electron are represented by a vector $\mathbf{v_0}=(dx,\theta_x,dy,\theta_y,dl,dp/p_0)$ where all components of $\mathbf{v_0}$ are taken with respect to the central trajectory $\mathbf{v_0}=0$. For instance, $dp/p_0$ is the relative momentum displacement of the electron and $dl$ is the path length difference with respect to the reference electron. One can obtain information on the temporal displacement of electrons by writing that $dt = dl/v$ where $v$ is the electron velocity. Transport through the beamline is described by equation $\mathbf{v}=M\mathbf{v_0}$ where $M$ is a $6\times6$ matrix representing all magnetic elements.

One can use this model to estimate the effect of the bend magnet on temporal compression. Let us first consider only longitudinal dynamics in the bend magnet, i.e. $dx\ll1$ and $\theta_x\ll1$. From the bend magnet matrix, one can easily show that a bunch is longitudinally recompressed after a drift corresponding to the temporal focal length of the bend magnet:
\begin{equation}
f_{bend}=(\alpha-\sin\alpha)\frac{mc}{eB_0}\gamma^3
\end{equation}
where $e$ is the electron charge, $\alpha$ the bending angle and $B_0$ the bending magnetic field. With the parameters of our beamline  ($\alpha\simeq45^{\circ}$ and $B_0=0.35$~T), this implies that the bend magnet can compensate a drift of $L_{d}=f_{bend}=45$~cm, which defines the total length of the beamline.

 \begin{figure}[t!]
\centerline{\includegraphics[width=8cm]{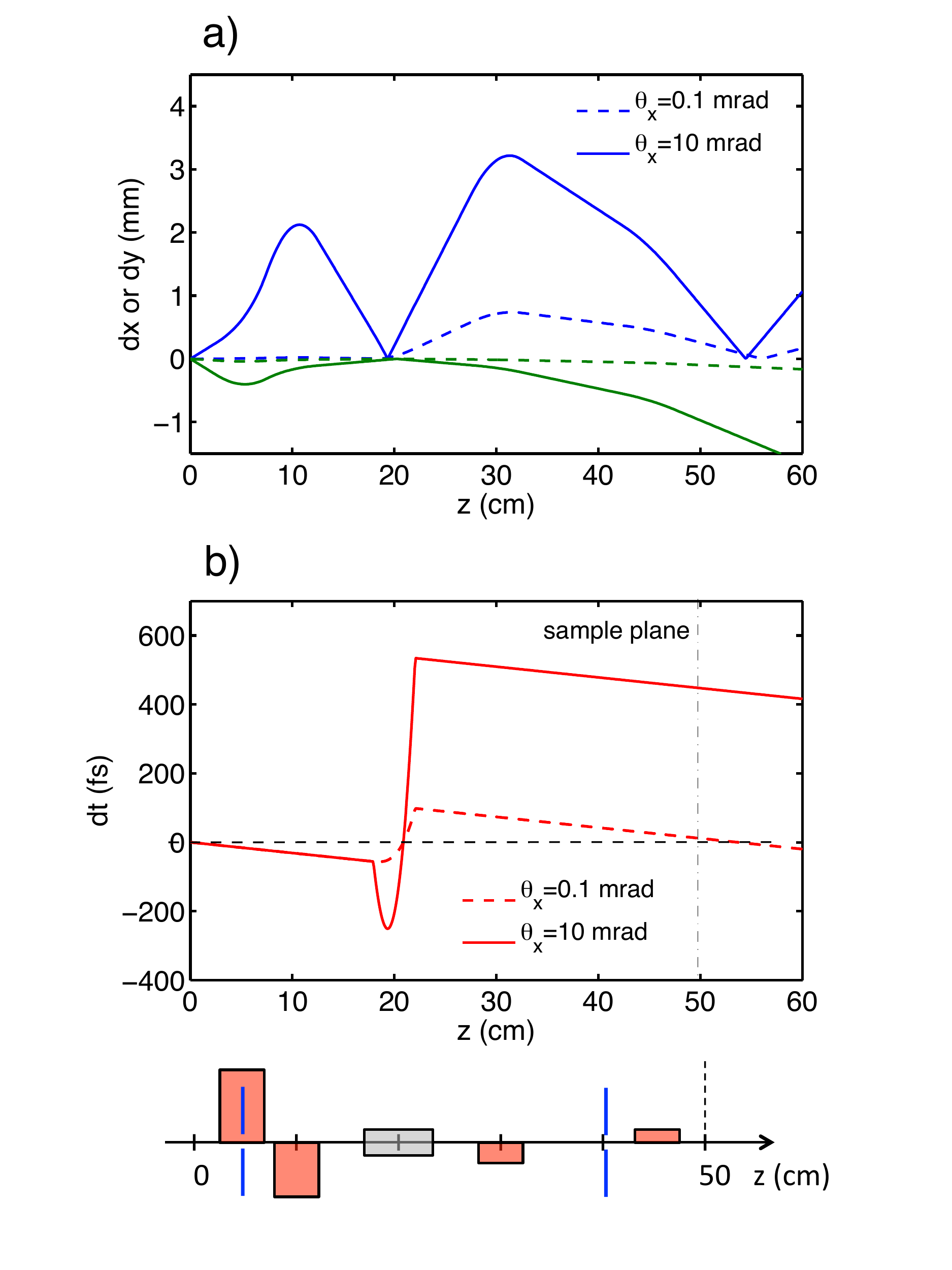}}
\caption{Ray tracing model showing electron trajectories through the beamline. a) Blue lines (green line): relative displacement $dx$ ($dy$) with respect to the central trajectory. These trajectories originate from the axis $dx=dy=0$ and have initial divergences $\theta_x=\theta_y=10$~mrad (full line) and $\theta_x=\theta_y=0.1$~mrad (dashed line), their momentum is shifted $dp/p_0=1\%$. b) Relative time displacement $dt\simeq dl/c$ for the same trajectories. Bottom: schematic of the beamline.}\label{fig3}
\end{figure}

This analysis does not hold when considering transverse effects: a finite $dx$ and a finite angle $\theta_x$ also contribute to the temporal shift of the trajectory. In this case the temporal shift in the bend $dt$ reads
\begin{equation}
vdt=dx\sin\alpha +(1-\cos\alpha)\rho_0\theta_x+(\alpha-\sin\alpha)\rho_0\frac{dp}{p_0}
\end{equation}
where $\rho_0=p_0/eB_0$ is the radius of curvature of the central trajectory in the bend. When these transverse terms are large, the bend does not recompress the bunch correctly. These transverse effects can be neglected when 
\begin{eqnarray}
dx & \ll & \frac{\alpha-\sin\alpha}{\sin\alpha}\rho_0\frac{dp}{p_0} \\
\theta_x & \ll & \frac{\alpha-\sin\alpha}{1-\cos\alpha}\frac{dp}{p_0}
\end{eqnarray}
With our beamline parameters, this means that transverse effects in the bend magnet can be neglected if $dx\ll50\mic$ and $\theta_x\ll2\mrad$. This explains why large divergence angles are not acceptable to achieve recompression. Our beamline design ensures that $dx$ is small by focusing the beam inside the bend magnet and the reduction of the divergence is achieved by spatial filtering with the two slits.

These effects are illustrated in fig.~\ref{fig3} which shows the the results of a standard ray tracing model. Figure ~\ref{fig3}a) shows the trajectory of electrons with $\theta=0.1\mrad$ and $\theta=10\mrad$ through the beamline. The effect on compression is quite dramatic, as can be seen in Fig~\ref{fig3}b): for the low divergence case, transverse effects are negligible and the bend recompresses the bunch ($dt=0$) at the sample plane, $z=f_{bend}=50$~cm. This is not the case for a large divergence: in this case, the duration reaches $\sim 500\fs$ in the sample plane. We conclude that the beamline is suitable for recompressing the electron bunch provided that slits are used to trim the angular distribution, which comes at the expense of a large loss of beam charge.

\section{Numerical simulations of the beam transport}\label{secGPT}
The optimization of this beamline is complex because it needs to take into account multiple and potentially conflicting objectives. Indeed, the beamline has to deliver at the sample plane (i) a narrow energy spread beam (ii) a small beam size $\sim 100\mic$ (iii) a large coherence length $>1$~nm, (iv) the shortest bunch duration  and (v) the largest amount of charge. 

To solve this problem, we used a genetic algorithm in the GPT code to find the optimum parameters of the beamline. The genetic algorithm assigns random values to a set of free parameters in the beamline, namely the quadrupole field strength, the size of the slits and the bend angle. First, a large number of beamline configurations are generated randomly in this manner. GPT is then used to propagate the electron beam through all configurations and to estimate the beam parameters in the sample plane, computing a few figures of merit for assessing the beam quality (bunch duration, coherence length, charge). The results are then ranked according to these figures of merit, the best results are selected for producing new configurations and the algorithm proceeds to a new iteration. We used populations of 300 configurations and found that the algorithm converges in less than 100 iterations. 

Concerning beam transport modeled by GPT, all magnetic elements have realistic geometries and include fringe fields \cite{mura14}. Space charge is not taken into account for the optimization with the genetic algorithm in order to keep computational time to a reasonable level. For the optimization, we used a simplified model distribution which approaches the electron distribution given by the PIC simulations as described in section \ref{secPIC} - see also table~\ref{table1} for all details on this model distribution. For this distribution, we assume that the initial bunch has zero duration so that the output duration at the sample plane is given by the response function of the beamline. Finally, unlike the output of the PIC simulations, we assume that the distribution has a narrow energy spread of a few percent. This is justified by the fact that the beamline is an excellent energy filter and electrons with energies outside of a narrow range are not transmitted, and therefore do not need to be taken into account. This also speeds up the computational time considerably.

\begin{figure}[t!]
\centerline{\includegraphics[width=9cm]{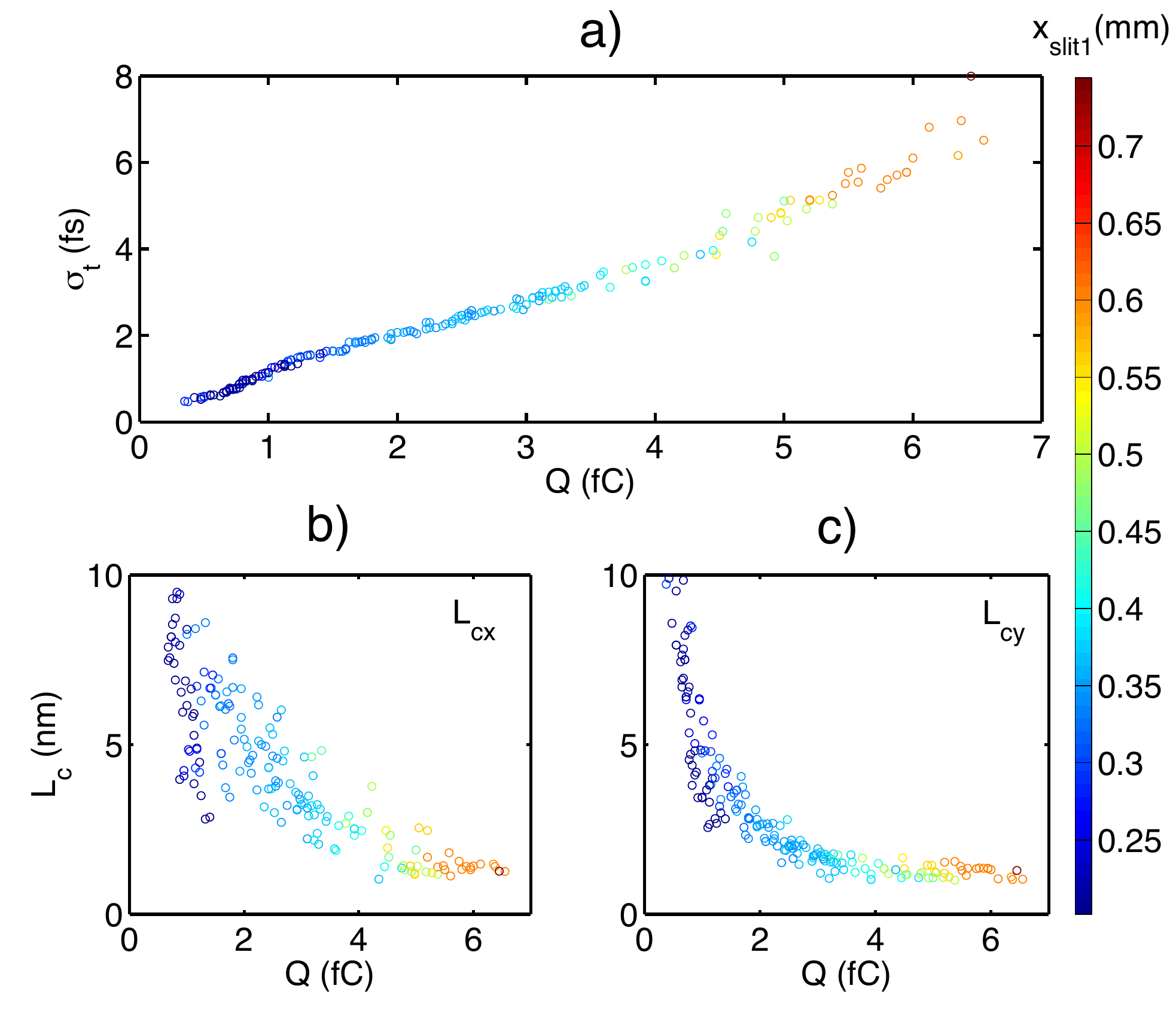}}
\caption{Results of genetic optimization using GPT code. a) Optimized configurations after convergence of the algorithm showing the r.m.s. duration of the bunch $\sigma_t$ versus its charge $Q$. b) and c) show the coherence lengths $L_{cx}$ and $L_{cy}$ versus charge for optimized configurations of the beam line. The color scheme represent the opening along $x$ of the first slit in the beamline.}\label{fig4}
\end{figure}

Figure~\ref{fig4} shows the results of a genetic optimization of the beamline. In fig.~\ref{fig4}a, each point represents a beamline configuration in a plot of r.m.s. duration $\sigma_t$ versus charge. This plot shows that fC bunches with sub-10~fs duration can be obtained, which is a remarkable result. Indeed, fC charge at kHz repetition rate permits to obtain high quality diffraction images in seconds or less. From fig.~\ref{fig4}a, it is also clear that there is a compromise between the charge that the beamline lets through and the bunch duration. In fig.~\ref{fig4}, the colorbar represents the opening of the first slit along the $x$-direction, and fig.~\ref{fig4}a shows that by opening the slit, the final beam has more charge at the expense of a longer duration. This indicates that this slit can be used as a knob for the beam charge and its duration. Indeed, except for the slit sizes, the parameters of the beamline operate in a narrow range once the genetic algorithm has converged: the quadrupole strength or bend angle vary by less than $1\%$ for all beamline configurations. Fig.~\ref{fig4}b) and \ref{fig4}c) show the coherence lengths, respectively $L_{cx}$ and $L_{cy}$ versus charge for various optimized beamlines. The coherence length reaches a few nanometers in all cases. Here again, beams with a smaller charge lead to a higher coherence length and the slit size appears to provide some level of tunability.  

Note that the beamline is composed of realistic off-the-shelf magnets, for example, the gradient in the quadrupoles are all $<20\;\mathrm{T/m}$ and the magnetic field of the bending magnet is modest, 0.35~T.

\begin{figure}
\centerline{\includegraphics[width=8cm]{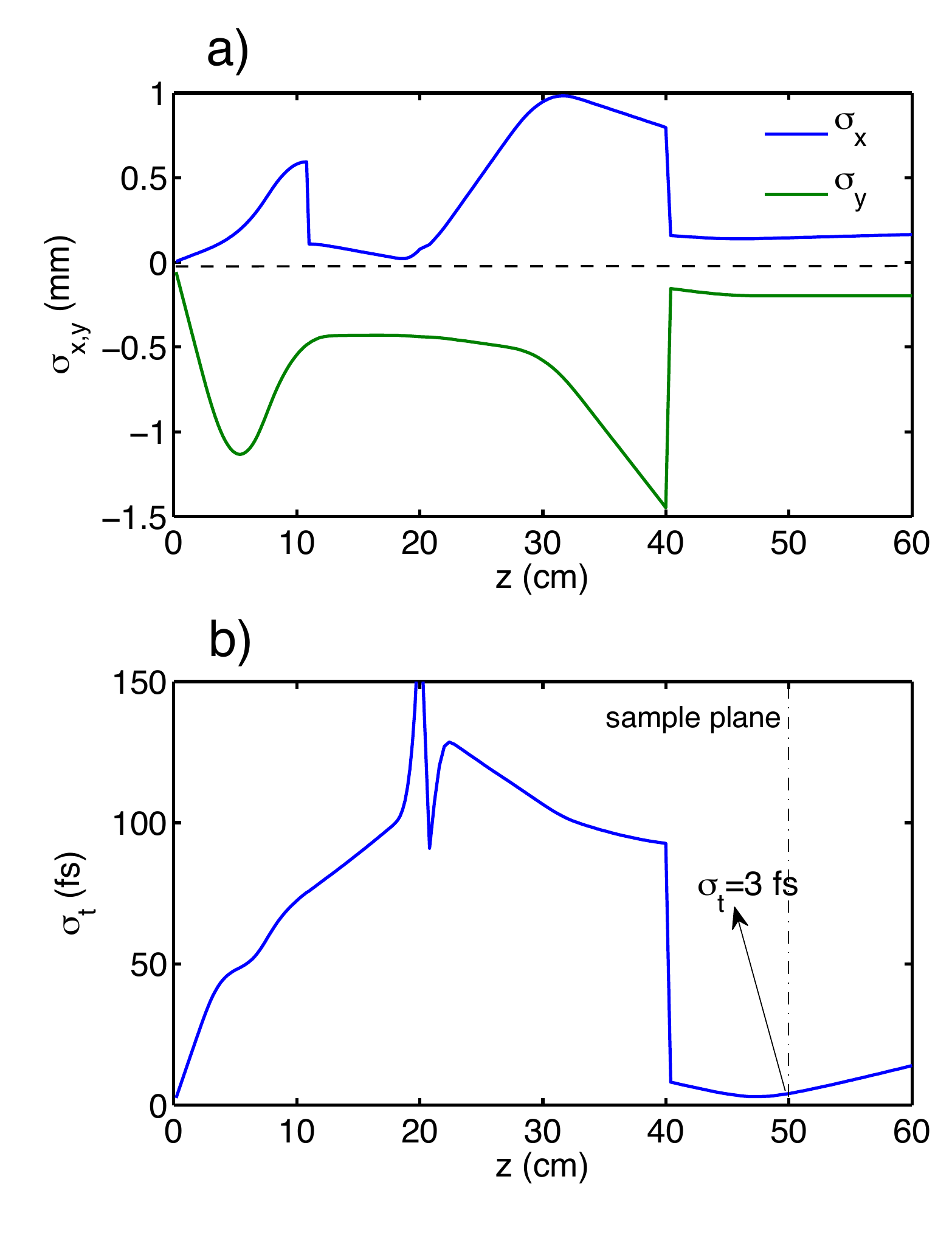}}
\caption{Results of GPT simulations with the model distribution (see Table 1). a) r.m.s. transverse beam sizes of the beam enveloppe through the beamline: $\sigma_x$ (blue line) and $\sigma_y$ (green line). b) r.m.s. bunch duration 
$\sigma_t$ through the beamline. The discontinuity at $z=20$~cm is an artifact of the r.m.s. calculation due to a change of reference frame during propagation in the bend magnet. }\label{fig5}
\end{figure}

\begin{table*}[htdp]
\scriptsize
\centering
\begin{tabular}{|c|c|c|c|c|c|c|c|c|c|c|c|}
\hline
\textbf{distribution} & Charge (fC) & $\;\;\gamma_0\;\;$ & $\sigma_\gamma/\gamma_0$(\%) & $\sigma_x$($\mic$) & $\sigma_y$($\mic$) & $\sigma_{p_x}(mc) $ & $\sigma_{p_y}$(mc)  & $\varepsilon_{nx}$(nm) &$\varepsilon_{nx}$(nm) & $L_{cx}$(nm) & $L_{cy}$(nm)\\
\hline
Model & 500 & 10.3 & 4.8 & 0.2 & 1 & 0.1 & 1 & 1.6 & 83 &- & - \\
(uniform) &  &  &  &  &  &  &  &  &  & &  \\
\hline
At sample plane ) & 2.8 & 10.3 & 0.3 & 150 & 200 & $2\times10^{-3}$ & $2\times10^{-4}$ & 15 & 40 & 4 & 2 \\
(from model) &  & &  &  &  &  &  &  & &  &  \\
\hline
Input distribution & 770 & 10.3 & 1.2 & 0.1 & 0.6 & 0.1 & 0.6 & 8 & 300 & -  & - \\
from PIC & &  &  &  &  &  &  &  &  &   &  \\
\hline
At sample plane  & 1.2 & 10.3 & 0.3 & 145 & 200 &  $2\times10^{-3}$ &  $2\times10^{-4}$ & 15 & 50 & 4 & 2 \\
(from PIC) &  &  &  &  &  &  &  &  &  &  &  \\
\hline
\end{tabular}
\caption{Parameters of the various distributions used in this work. Row~1: a distribution that models the PIC distribution showed in fig.~\ref{fig1}. The distribution is uniform, thus the $\sigma$ values correspond to the width of uniform distributions. Row~2, parameters of the output distribution through the beamline when the model distribution is used (as in fig.~\ref{fig4} and fig.~\ref{fig5}). Row~3: parameters of the distribution from the PIC simulations shown fig.~\ref{fig1}. Note that the energy spread was trimmed to $2\%$ at FHWM for making computing in the particle tracking simulations easier. Row~4: the resulting distribution at the sample plane obtained when the PIC distribution is used. For row~2 to row~4, the various $\sigma$ represent the r.m.s. values of the distributions.}
\label{table1}
\end{table*}%
We now focus on a single beamline configuration in order to gain insight on the beam dynamics during transport. Figure~\ref{fig5}a) shows the r.m.s. beam size through the beamline. Note in particular the effect of the first slit, which in conjunction with the first two quadrupoles deliver a near collimated beam inside the bend magnet, with a small transverse size along $x$. This reduces transverse effects in the bend magnet and permit longitudinal compression of the bunch. Longitudinal dynamics is plotted in fig.~\ref{fig5}b) where the blue line represents the r.m.s. bunch duration. The second slit trims the distribution even more and eliminates electrons with larger angles and traverse positions. This results in compression to sub-10~fs in the sample plane and a beam size in the $100\mic$ range. Without the second slit, the bunch duration would be in the $~100\fs$ range so that a gain of one order of magnitude is achieved at the expense of a loss of charge by more than an order of magnitude.

\begin{figure}
\centerline{\includegraphics[width=8cm]{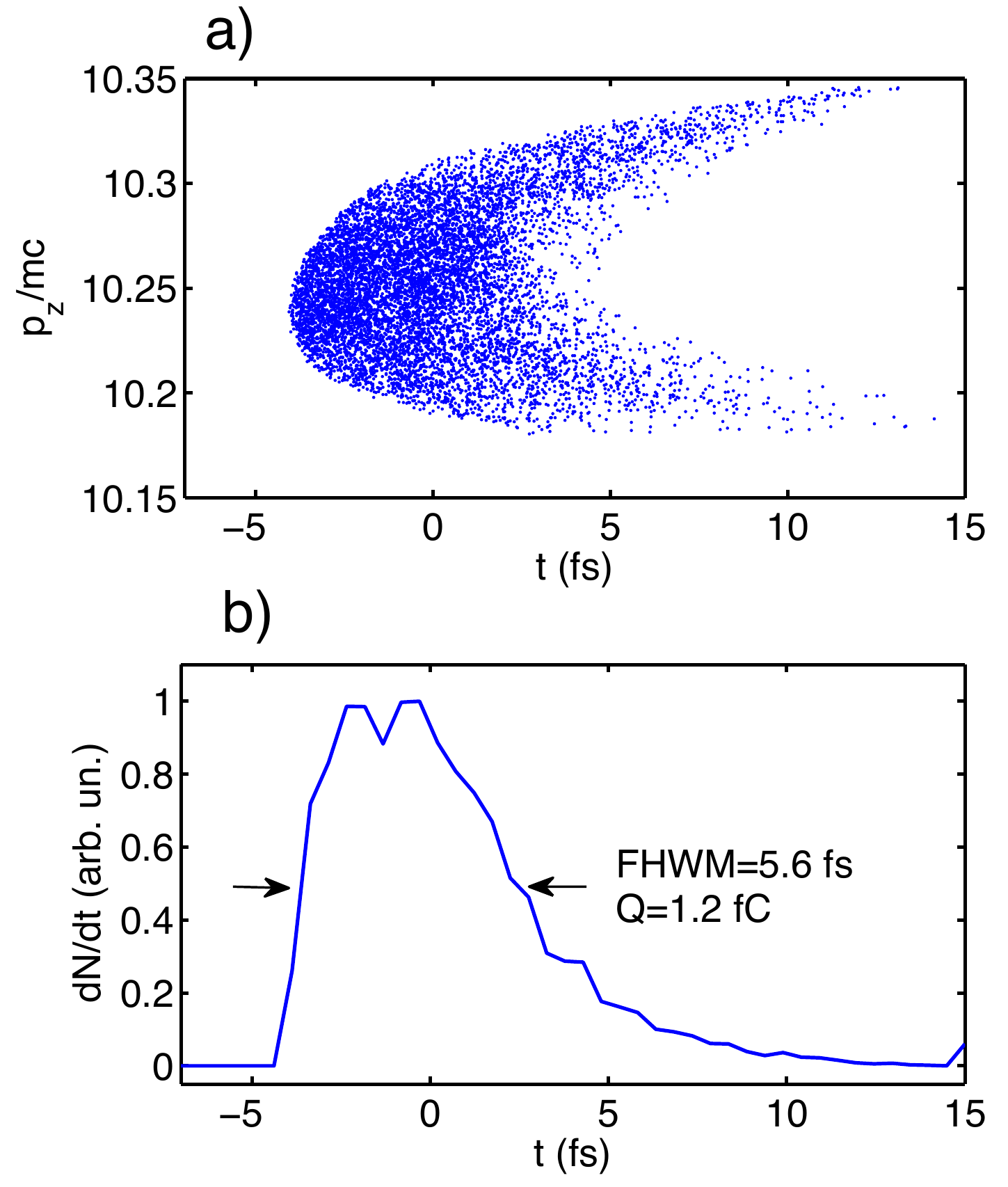}}
\caption{Results of GPT simulations using the electron distribution from the PIC simulations. a) Electron distribution in phase space $(p_z,t)$ for the compressed bunch in the sample plane. b) Temporal distribution of the bunch containing a charge of $1.2$~fC. The bunch is $5.6$~fs at FWHM. }\label{fig6}
\end{figure}

Finally, we checked that the compression beamline also works for a realistic distribution. Therefore, we used the distribution from the PIC simulations as an input for GPT for the same beamline configuration used in fig.~\ref{fig5}. As mentioned earlier, the distribution from the PIC simulation was trimmed in energy down to 2\% FWHM in order to speed up the computational time of the particle tracking simulations. For this simulation, space charge was taken into account throughout the whole propagation \cite{popl04}. Therefore, the whole beamline, from the generation of the electron beam to its propagation and transport, is simulated numerically without any extra assumptions. Figure~\ref{fig6} confirms that the compression concept is still valid when using the real PIC distribution: fig.~\ref{fig6}a) shows the particle distribution in longitudinal phase space whereas fig.~\ref{fig6}b) shows the temporal distribution of the electron bunch, showing that a bunch of 5.6~fs at FWHM and containing 1.2 fC can be achieved. We found that space charge has a minor impact on the results, apart from decreasing the transmitted charge by about $10\%$.

All the parameters of the input beam distribution are given in Table 1. The table also shows the parameters of the distribution in the sample plane. It is clear that the distribution in the sample plane is relatively independent of the input, showing that the beamline effectively trims the beam and provides the target parameters in terms of beam size and coherence length. The sub 10-fs pulse duration in combination with a fC charge on a $\sim100\mic$ spot with a transverse coherence length of 4 nm makes the electron source very relevant for use in ultrafast electron diffraction experiments. The laser-based solution introduces no rf-jitter, and the table-top design with just a few off-the-shelve components makes this a very practical and cost-effective solution for scientific applications. 

Finally, the transmission of the beamline is less than $1\%$ and provides fC charge, which is enough for applications. However, the transmission could be further improved by decreasing the divergence angle at the source which might be possible using plasma lensing techniques \cite{thaury15} or very strong quadrupoles \cite{eich07}. In this case, charge of tens of fC might become available.

\section{Conclusion}

In conclusion, using numerical simulations, we have demonstrated that a 5 MeV electron beam can be generated in a laser-plasma accelerator driven by a kHz class laser system. In the plasma, the electron bunches have durations of a few femtoseconds. Subsequently we have shown a simple but effective beamline that is capable of a) trimming undesired parts of phase space, while b) recompressing the remaining part of the beam temporaly, while c) focusing the beam on a target such that it is useful for ultrafast electron diffraction experiments. The main advantages of the proposed laser-based approach are ultra-short pulse-duration, elimination of rf-jitter, simplicity of the beamline and quality of the beam (just don’t mention total charge).

\begin{acknowledgments}
This work was funded by the European Research Council under Contract No. 306708, ERC Starting Grant FEMTOELEC 
\end{acknowledgments}


\end{document}